\documentclass[12pt,preprint]{emulateapj}
\usepackage{ulem}
\usepackage{lscape}

\newcommand{\HI}{\ion{H}{1}} 
\newcommand{\CIV}{\ion{C}{4}}
\newcommand{\CIII}{\ion{C}{3}}
\newcommand{\CII}{\ion{C}{2}}
\newcommand{\OVI}{\ion{O}{6}}
\newcommand{\OI}{\ion{O}{1}}
\newcommand{\NV}{\ion{N}{5}}
\newcommand{\MgII}{\ion{Mg}{2}}
\newcommand{\SiIV}{\ion{Si}{4}}
\newcommand{\SiIII}{\ion{Si}{3}}
\newcommand{\SiII}{\ion{Si}{2}}
\newcommand{\AlII}{\ion{Al}{2}}
\newcommand{\FeII}{\ion{Fe}{2}}
\newcommand{\cm}{{\rm cm}}
\newcommand{\kms}{\,{\rm km}\,{\rm s}^{-1}}
\newcommand{\lya}{Ly$\alpha$}
\newcommand{\lyb}{Ly$\beta$}
\newcommand{\lyg}{Ly$\gamma$}
\newcommand{\K}{{\,{\rm K}}}

\begin{document}

\title{THE COMPOSITE SPECTRUM OF STRONG LYMAN-ALPHA FOREST ABSORBERS}

\author{Matthew M. Pieri\altaffilmark{1,2}, Stephan Frank\altaffilmark{3}, David H. 
Weinberg\altaffilmark{1,4},  Smita Mathur\altaffilmark{1},\\ and Donald G. York\altaffilmark{5,6}}
\shortauthors{Pieri et al.}
\altaffiltext{1}{Department of Astronomy, The Ohio State University, 140 West 18th Avenue, Columbus, OH 43210, USA}
\altaffiltext{2}{CASA, Department of Astrophysical and Planetary Sciences, University of Colorado, 389 UCB, Boulder, CO 80309, USA; mpieri@colorado.edu}
\altaffiltext{3}{Observatoire Astronomique de Marseille-Provence, P\^ole de l'\'Etoile Site de Ch\^teau-Gombert, 38, rue Fr\'ed\'eric Joliot-Curie
13388 Marseille cedex 13, France}
\altaffiltext{4}{Institute for Advanced Study, Einstein Drive, Princeton, NJ 08540}
\altaffiltext{5}{Department of Astronomy and Astrophysics, University of Chicago, Chicago, IL 60637, USA}
\altaffiltext{6}{Enrico Fermi Institute, University of Chicago, Chicago, IL 60637, USA}

\begin{abstract}
We present a new method for probing the physical conditions and metal enrichment
of the Intergalactic Medium: the composite spectrum of \lya\
forest absorbers. We apply this technique to a sample of 9480 \lya\ absorbers
with redshift $2 < z < 3.5$ identified in the spectra of 13,279 high-redshift
quasars from the Sloan Digital Sky Survey (SDSS) Fifth Data Release (DR5).
Absorbers are selected as local minima in the spectra with 
$2.4 < \tau_{\rm Ly\alpha} < 4.0$; at SDSS resolution ($\approx 150\kms$ FWHM), these
absorbers are blends of systems that are individually weaker. In the
stacked spectra we detect seven Lyman-series lines and metal lines of 
\OVI, \NV, \CIV, \CIII, \SiIV, \CII, \AlII, \SiII, \FeII, \MgII, and \OI. Many of these 
lines have peak optical depths of $<0.02$, but they are nonetheless detected
at high statistical significance.  
Modeling the Lyman-series measurements implies that our 
selected systems have total \HI\ column densities 
$N_{\rm HI} \approx 10^{15.4}\cm^{-2}$. Assuming typical physical conditions 
$\rho/\bar{\rho}=10$, $T=10^4-10^{4.5}\K$, and [Fe/H]$=-2$ yields reasonable 
agreement with the line strengths of high-ionization species, but it 
underpredicts the low-ionization species by two orders of magnitude or more. 
This discrepancy suggests that the low ionization lines arise in dense, cool, 
metal-rich clumps, present in some absorption systems.

\end{abstract}


\section{INTRODUCTION}

The Lyman-$\alpha$ forest is a tracer of the diffuse matter between galaxies known as 
the Intergalactic Medium (IGM), and at high redshift ($2<z<4$) this diffuse medium
contains roughly 70\% of the baryonic mass and occupies 90\% or more of the volume.
Galaxies form from the collapse of dark matter structures and corresponding 
accretion of gas from the IGM.  But this is not a one-way process, and feedback is vital for a 
complete picture. Galactic outflows modify the kinetic and thermal energy of the medium 
and distribute the by-product of star formation: metals (e.g. \citealt{mf99, mfr01, od06, 
pmg07, pm07}). 

Metals have been observed in the high redshift \lya\ forest. In particular \CIV\ (e.g. 
\citealt{my87, c95,sc96, e00, s03, psa06}) and \OVI\ (e.g. \citealt{s00, ph04,ssr04, bh05, 
a08, f10, p10}) provide prominent absorbers and are commonly measured. Other species 
seen are \SiIV, \SiIII, \CIII\ and \NV\  (e.g. \citealt{s03, a04, fr09}). These measurements 
provide some indications of the regions of the Universe touched by mechanical feedback, 
but they can also provide a useful probe of the ionization properties of the medium (and 
so the physical conditions of the gas), and the abundance pattern. The main obstacle to 
significant progress on the latter two goals is a lack of large numbers of metal species to 
measure both from the same element (to measure ionization characteristics) and from 
different elements (to measure abundance patterns).

In this paper, we introduce a new 
technique for the detection of weak metal lines --- the composite spectrum of \lya\ forest
absorbers.
We apply this approach to the largest \lya\ forest dataset available: the Sloan Digital Sky 
Survey (SDSS) sample of QSO spectra. \citet{p10} demonstrated that the SDSS sample 
can provide precision measurements of weak lines by searching for 
\OVI\ absorption associated with the bulk of the \lya\ forest.
Here we produce a blind search for absorption correlated with the strong \lya\ forest absorbers;
we identify and fit 19 metal lines.

\section{Production of Composite Spectra}
\label{method}

\subsection{The Sample and Lyman-$\alpha$ Forest Line Selection}
\label{selection}

We use the Sloan Digital Sky Survey Data Release 5 (SDSS DR5) \citep{aetal07}, which 
provides 13,279 QSOs with useful forest coverage in the redshift range  $2<z<3.5$. The 
spectra are taken from the QSO Absorption Line Sample 
(QSOALS; \citealt{y06}) in the same manner described in 
\citet{p10} employed for DR3 spectra. The spectral resolution is wavelength dependent 
and varies from $R=1800 - 2200$. In the following work we assume that $R=2000$ at all 
wavelengths, and the error introduced by this approximation is smaller than the quoted 
errors.

We define ``absorbers'' in a simple way,  selecting
pixels that are lower in flux than their two neighboring pixels
and have optical depth $2.4<\tau_{\rm Ly\alpha}<4$. 
We require that our \lya\ absorbers are redward of the \lyb\ forest and blueward of $5000~
\kms$ from the quasar \lya\ emission redshift.  In the stacked spectrum discussed in this paper we 
use only \lya\ absorbers in the redshift range $2<z<3.5$, resulting in some variation in the 
average redshift across the stacked spectrum (see Figure~\ref{stack}). 
We ensure that the selected absorbers are not saturated at SDSS resolution
by requiring a minimum flux of $\sigma_n/2$.
This requirement, which eliminates damped \lya\ systems and noisy 
lines, discards 
86\% of \lya\ absorbers in the required optical depth range. 
At SDSS spectral resolution, solitary
\lya\ forest lines with $b-$parameters $\sim 30\kms$ and no damping wings
do not reach an apparent optical depth 
greater than {$\tau_{\rm Ly\alpha} \approx 1.9$}, even if they
are fully saturated.
The absorption features in our sample are therefore blends of several
\lya\ forest lines rather than individual strong lines; we address
the typical column densities of these features in
\S \ref{interpret} below,
using the stacked spectra themselves.

\subsection{Spectral Stacking Methodology}

For each \lya\ absorber in our sample, 
we de-redshift the whole spectrum to its rest-frame.
Each quasar spectrum is carried forward once for every \lya\ absorber, so some 
SDSS spectra are used more than once while others are not used at all.  What results is a 
stack of spectra on a rest-frame wavelength grid. We employ two statistics to produce 
composite spectra: the median, and the arithmetic mean with a 3\% outlier 
clipping. We require a minimum of 100 pixels for the measurement of the stacked 
spectrum at any point.  When computing the composite spectrum redward of 1236\AA, we only 
include pixels if they are outside of the \lya\ forest of the quasar spectrum in question,
since the pixels outside of the forest have much less background absorption and 
associated random fluctuations.

We estimate the error in the flux for the stacked spectrum by bootstrapping the data using 
100 realizations. 
This is mostly consistent with the estimate from the propagation of errors, but it is larger in the wavelength range between the \lyb\ and \lya\ lines, due to scatter in uncorrelated \lya\ forest absorption. 

\begin{figure}[htbp]
\begin{center}
\mbox{
\includegraphics[angle=90,width=.97\columnwidth]{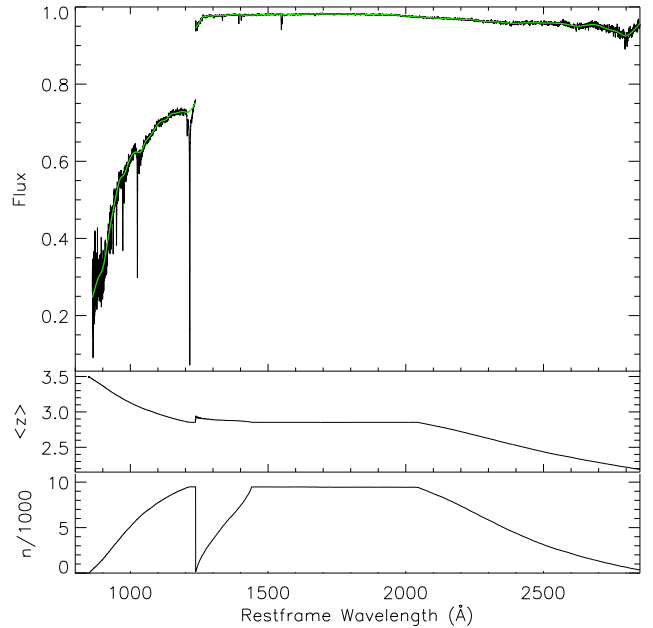}}
\caption{{\it Top panel}: The median stacked spectrum of 9480 \lya\ absorbers. Overlaid in 
green is the `pseudo-continuum' set by uncorrelated absorption. Lyman series lines and 
other lines can be seen along with a broad absorption trough caused by uncorrelated Lyman 
series lines. {\it Middle panel}: The mean redshift of pixels that make up the stacked 
spectrum. {\it Bottom panel}: The number of pixels that go into the stacked spectrum at 
every wavelength.  
}
\label{stack}
\end{center}
\end{figure}

Figure \ref{stack} shows the stacked spectrum we obtain using the median (which is our 
fiducial choice).  
The stacked spectrum never reaches 
100\% transmission ($F=1$) 
because there is always some 
uncorrelated, contaminating absorption.

\subsection{Composite Spectra of Lyman-$\alpha$ Forest Absorbers}
\label{makecomposite}

The full-width-half-maximum of SDSS resolution is $136-167~\kms$, which corresponds 
to an effective Doppler parameter of $b_{\rm SDSS}=82-100~\kms$. This is around four times 
as broad as typical Lyman-series lines and broader still for metal lines. Hence the 
majority of contaminating ``background'' absorption in our spectra is not a true superposition of 
overlapping lines but a blending of distinct lines that appear 
superimposed due to SDSS resolution. 

Given that uncorrelated contaminating absorption varies smoothly in our stacked 
spectrum, we can treat it as a continuum by performing a standard spline fit,
producing the `pseudo-continuum' shown by the
overlay in Figure \ref{stack}.  We have scaled away its effect by adding the 
flux decrement of the pseudo-continuum to our stacked spectrum, thus arriving at what 
is effectively a composite rest-frame spectrum of the selected \lya\ absorbers. 

\begin{figure*}
\begin{center}
\mbox{
\includegraphics[angle=0,width=0.8\textwidth]{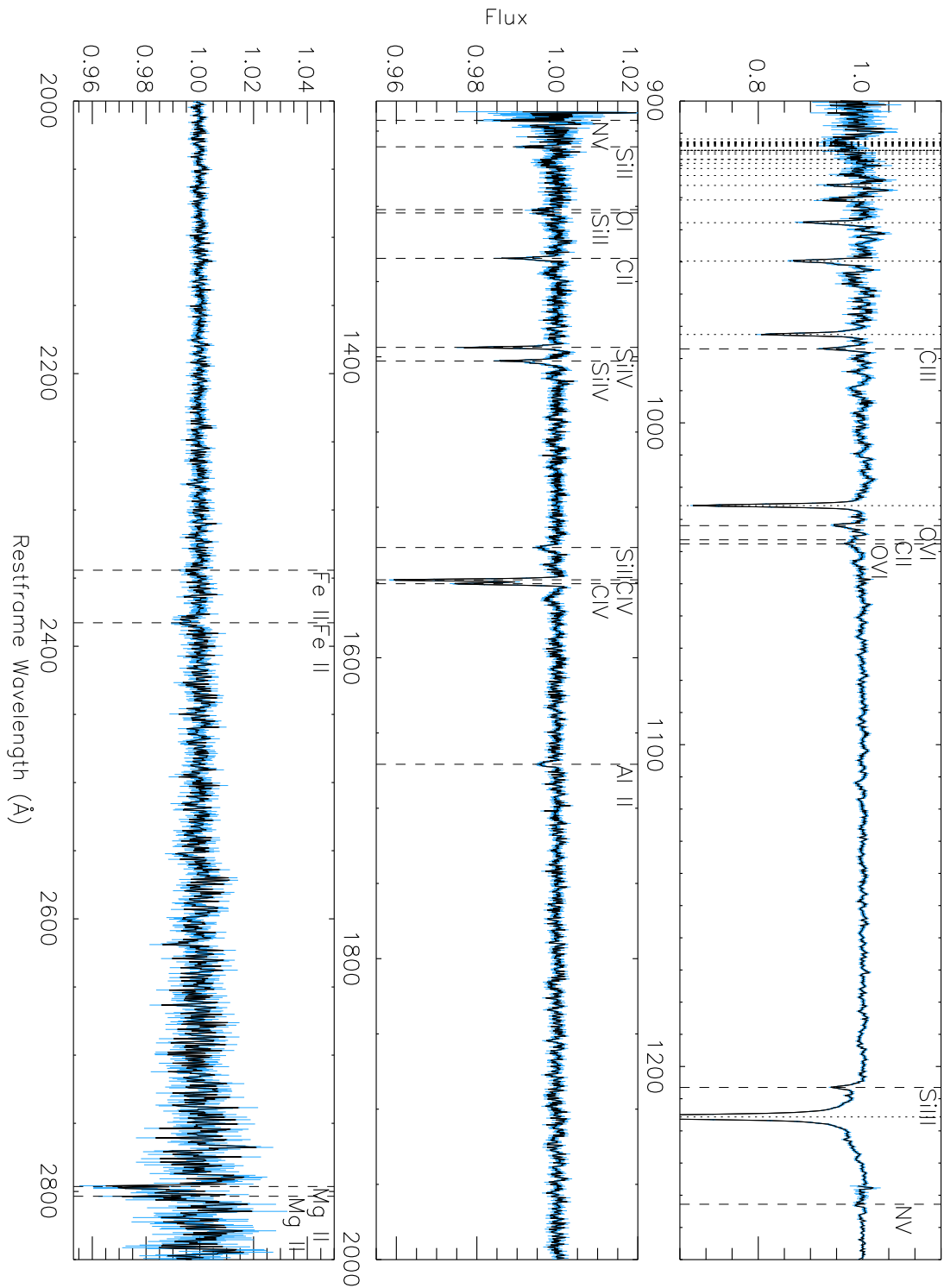}
}
\caption{ The rest-frame spectrum of our selected \lya\ absorbers. The blue vertical lines 
indicate the error estimate.  This composite is produced by scaling away the `pseudo-
continuum' from the median stacked spectrum shown in Figure \ref{stack}. Lyman series 
lines are marked ({\it dotted lines}). Metal lines found at high confidence are marked ({\it 
dashed lines}) with labels to the right. These metal lines have been fitted and the results 
of these fits are shown in Table~\ref{linefits}. Note the difference in scales for both axes 
for all panels.}
\label{compspec}
\end{center}
\end{figure*}

Figure \ref{compspec} shows the median spectral stack of our sample of \lya\ forest 
absorbers renormalized using the pseudo-continuum. 
A great many absorption lines are seen in this composite rest-frame 
spectrum. The Lyman series is particularly clear and is marked by dotted lines. Numerous 
metal lines are also seen at high confidence and are marked with dashed lines and 
labelled. 

The \lya\ line shows large wings, which are to be expected as a signal of large scale 
structure. The scale of the correlation extends to $\sim 3000\kms$, which is in 
good agreement with the findings of \citet{m06}. There is no indication that the signal 
arises from damping wings, and higher order Lyman lines rule out this interpretation (see 
following section). These wings have not been carefully fitted to separate the signal of 
clustering from the mean flux decrement in the forest, as the measurements in this paper 
are not dependent on this fit.

It should be noted that the weak lines seen redward of some lines  (e.g. \OVI, \SiIV, and 
\CIV\ doublets) are a signal due to \SiIII\ interlopers. There is a small sample of strong 
\SiIII\ absorbers that has entered our sample of \lya\ lines for stacking. When these lines 
are misinterpreted as \lya, a ``shadow'' signal in our stacked spectra is seen shifted by $
\lambda_{{\rm Ly}\alpha}/\lambda_{{\rm SiIII}}=1216/1206$. This shadowing is apparent for every 
identified metal line in the arithmetic mean stack, in keeping with its status as a less 
outlier-resistant statistic. 
We disregard these ``shadow'' lines in our analysis below.
It is surprising that no substantial shadow is seen for \lya; we do not 
have a simple explanation for this effect, and it appears that these \SiIII\ absorbers are an 
interesting sample in their own right.

\section{Line Fitting and Interpretation}
\label{interpret}

Full interpretation of these stacked spectra will require the use of model
spectra and/or degraded high-resolution spectra to account for the effects
of line blending, sample selection, and pseudo-continuum subtraction.
In this paper, we restrict our interpretation to some general conclusions
that can be drawn by comparing the measured line strengths to simple models.

\subsection{ \HI\ Column Density}

\begin{figure}
\begin{center}
\begin{tabular}{c}
\includegraphics[angle=90,width=.95\columnwidth]{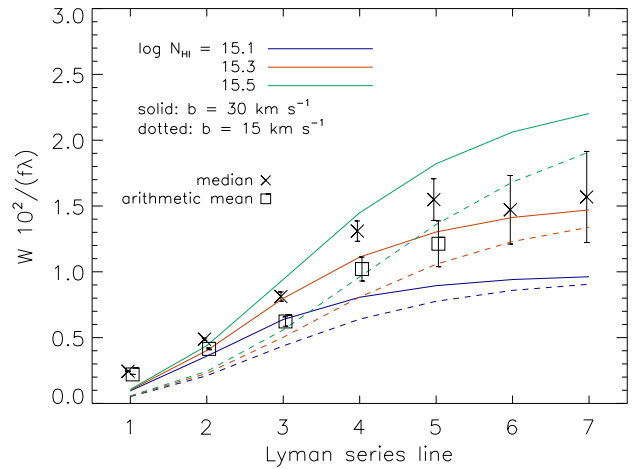} \\
\end{tabular}
\caption{Curve of growth for the first seven Lyman series lines,
labelled 1 (\lya) to 7 (Ly$\eta$). Points show the rest-frame
equivalent widths normalized by $f\lambda$ measured from the
median stack ({\it crosses}) and the arithmetic mean stack ({\it squares}).  
Curves show predictions for Gaussian lines with $b$-parameters
of $30\kms$ ({\it solid}) and $15\kms$ ({\it dashed}) and column densities
$\log N_{\rm HI}/\cm^{-2} =15.1$, 15.3, 15.5 (bottom to top).}
\label{CoG}
\end{center}
\end{figure}

Figure \ref{CoG} shows a measurement of the \HI\ column density
using the Lyman series. For each of seven Lyman series lines (n=1:\lya, 
n=2:\lyb, n=3:\lyg\ etc.), we measure a rest-frame equivalent width $W$.
We plot them normalized
by the product $f\lambda$
of oscillator strength and line wavelength.
Optically thin Lyman series lines would all have equal values of $W/f\lambda$.
The equivalent width is shown for both the median and the arithmetic mean, with error 
bars drawn from bootstrapping in the composite spectra.
The noise in the arithmetic mean spectrum renders the pseudo-continuum 
fitting blueward of 935\AA\ unreliable, so we only show Lyman lines n=1,2,3,4 and 5. 

The curves in Figure \ref{CoG} are created by using
VPFIT\footnote{http://www.ast.cam.ac.uk/ $\sim$rfc/vpfit.html} to generate 
model lines at SDSS resolution, from which we measure a simulated $W$.
The flatness of the $n=4-7$ points in the median stack suggests that these
lines are optically thin, and comparison to the model curves then implies
a total column density $\log N_{\rm HI}/\cm^{-2} \approx 15.3-15.5$
for our median absorption features.  In this column density range,
lines with velocity width $b=30\kms$ go from significantly 
saturated at $n=3$ to minimally saturated at $n=4$, in agreement with
the trend in the data points.  Lines with $b=15\kms$, corresponding
to the narrowest Doppler parameters seen for individual \lya\ forest 
absorbers, predict rising values of $W/f\lambda$ from $n=4-7$,
in clear disagreement with the observed trend.
Lines much broader than $b=30\kms$ would predict too much 
Ly$\beta$ and Ly$\gamma$ absorption relative to the higher order lines.

Of course, these absorption features are probably not well described
by single Voigt line profiles, in part because the composite
spectrum comes from systems with a range of properties, but mostly
because the contributing features are themselves blends of
multiple lines.  The models in Figure~\ref{CoG} all under-predict the
\lya\ equivalent width,
plausibly because the ``main'' absorption
feature is near-saturated in \lya\ and absorption outside the Gaussian
wings (but still within the SDSS resolution element)
makes a larger relative contribution.
While the $b$ parameters implied by Figure~\ref{CoG} have a complex and
non-trivial significance,
the high-order Lyman series lines do appear
to give a robust estimate of the total column density characteristic
of our median stack, with 
$\log N_{\rm HI}/\cm^{-2} \approx 15.3-15.5$.
Results for the arithmetic mean stack suggest a column density
lower by 0.1-0.2 dex, but it is harder to draw clear conclusions
because we cannot robustly measure Lyman-6 and Lyman-7
in the arithmetic mean stack.

\subsection{ Metal line Column Densities}

We have used VPFIT to fit all of the metal lines marked in Figure~\ref{compspec},
assuming spectral resolution $R=2000$.
Wherever possible, we measure each line independently. Only the blended transitions 
\ion{C}{2} with the weaker \OVI\ line, along with \ion{O}{1} and one of our \ion{Si}{2} lines 
($\lambda 1304$), are fitted with a joint analysis. 
Since contamination by \SiIII\ shadowing is a concern for the weaker \SiIV\ line we discard
it from the analysis.
Table~\ref{linefits} lists the rest wavelengths, equivalent widths ($W$) and the fitted column densities
($N$), Doppler parameters ($b$), and reduced $\chi^2$ values for
both the median and arithmetic mean stacks.
Many lines have fitted $b$-parameters significantly larger than the expected width
of metal lines. There is a large scatter in widths but low ionization lines are typically broader.
These large velocity widths could be a
signature of dispersion within the absorption systems or large-scale clustering
of the metal lines with the \lya\ absorbers (which is consistent with the \lya\ clustering signal). 
These results are in line with measurements of \CIV\ clustering on scales of up to $600~\kms$ \citep{psa06,s06}.

Regardless of the true line widths, these metal lines are likely to be optically
thin, making the derived column densities robust within the VPFIT-derived errors.
Multiple measurements of the same species for a given stack are usually consistent within 
quoted errors, with the notable exceptions of
\ion{Si}{2} and \CIV. In the case of  \ion{Si}{2}, the line at 1260\AA\  is discrepant with the 
other two measured lines. The continuum varies rapidly over a narrow wavelength range 
in this part of the spectrum, and we conclude that the continuum fitting for this line is the 
dominant source of error and therefore drop the line from further analysis.
The two lines of the \CIV\ doublet imply column densities that differ by 0.1 dex,
about five times the quoted 
$1\sigma$ errors. Also, the fit for the strongest doublet member is poor. Since this is
the strongest metal line, it may also have the strongest signal of clustering in 
the line profile, violating the single-line model assumption used in VPFIT.
We conclude that the precision of this measurement is at the 0.1 dex level.

\begin{figure}
\begin{center}
\begin{tabular}{c}
\includegraphics[angle=90,width=.95\columnwidth]{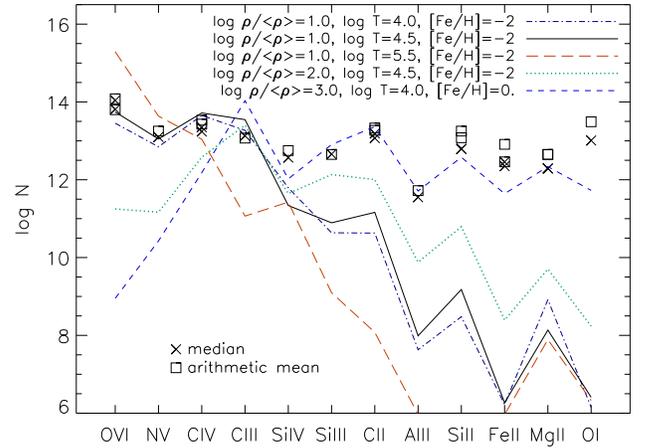} \\
\end{tabular}
\caption{The column densities of metal species measured in order of decreasing 
ionization potential. Column densities from our 
median and mean composite spectrum are shown as crosses 
and squares, respectively. Models curves are shown, assuming 
$log N_{\rm HI} =15.4$, a solar 
abundance pattern and a quasar+galaxy UV background. 
The $\rho/\bar{\rho}\sim 1000$ model curve is an example of clumpy conditions and the 
other curves represent densities and temperatures typical of the IGM as probed by the 
\lya\ forest at high redshift.}
\label{stackprojection}
\end{center}
\end{figure}

Figure~\ref{stackprojection} shows our column density measurements 
from the median and arithmetic mean stacks.  Species are listed in order of decreasing 
ionization potential.  The column densities 
derived from the mean stack tend to be slightly higher, by about 0.2 dex, 
but the overall agreement is good.
Somewhat counter-intuitively, the similarity of line strengths in the arithmetic mean and median
stacks does not imply that this level of absorption is present in most selected
absorbers. If the individual lines are not outliers with respect to the noise 
distribution, then a sub-population of absorbers can shift the mean and median flux 
decrements by similar amounts. By looking at the full distribution of flux decrements, one
can set limits on the fraction of systems that contribute most of the absorption. While we 
reserve detailed analysis to future work, our preliminary studies of the distribution of flux
decrements at line center suggests that the metal lines with higher ionization potentials 
(for \SiIII\ and greater) arise in 15--30\% of the selected \lya\ systems and that the lower
ionization lines (for \CII\  and lower) arise in 5--15\% of systems.

Despite this caveat about sample inhomogeneity,  it is useful
to compare our results with the simple predictions. Model curves are shown in Figure~\ref{stackprojection} using CLOUDY version 08.00 \citep{f98}, and assuming $\log N_{\rm HI} =15.4$, a solar abundance pattern and a quasar+galaxy UV background \citep{hm01}. Four curves 
show a range of models representing typical conditions in the \lya\ forest with a metallicity of 
[Fe/H]$=-2$.
Physical conditions $\rho/\bar{\rho}\approx 10$ and $T=10^{4}-10^{4.5}\K$ give
a reasonable match to the high-ionization lines,
but the strengths of the lower ionization lines are under-predicted by
two orders of magnitude or more. Therefore, lower ionization potential species are unlikely to be 
reproduced by models with diffuse IGM conditions. 
We can, however, get an approximate 
match to these lines with $\rho/\bar{\rho}\sim 1000$,
$T \approx 10^4\K$, and near-solar metallicities.
Since the \HI\ column densities implied by the Lyman-series lines are far
too low for self-shielding (which requires $N_{\rm HI} \ga 10^{17}\cm^{-2}$),
the absorbing gas should be photo-ionized by the UV background, so its
temperature is unlikely to be below $10^4\K$.\footnote{We cannot completely rule 
out the possibility that the low ionization species arise in a small fraction of systems that are  
self-shielded (Lyman Limit systems). It is unlikely that such a model could be reconciled with 
the low mean depth of Lyman-series lines, but this should be investigated further.}
The combination of these high densities with $N_{\rm HI} \approx 10^{15.4}\cm^{-2}$
implies $\sim 10$ pc pathlengths for
these low ionization absorbers. We conclude that different absorber populations,
or perhaps different gas phases within the same systems, are responsible for the high-
and low- ionization lines.

The high density, high metallicity, and short pathlength implied by this
analysis suggests that the low ionization metal lines may arise in
circumgalactic regions, perhaps in outflows from high redshift galaxies,
or perhaps in enriched filaments feeding these galaxies.
The \lya\ absorption features are similar to those
found in sightlines that pass within $\sim100~{\rm kpc}$ of Lyman break 
galaxies (Steidel et al. private communication; \citealt{s09}).

\tabletypesize{\normalsize}
\begin{deluxetable*}{l c c c c c c c c c c c} 
\tablecaption{Metal lines measured in the composite rest-frame spectrum of \lya\ 
absorbers}
\tablewidth{0pt}
\tablehead{
Species & $\lambda (\rm {\AA})$ & \multicolumn{4}{c} {Median} &&  \multicolumn{4}{c} 
{Arithmetic mean}\\
\cline{3-6} \cline{8-11}
&& $log N$  & $b (\kms)$ & $\chi^2_r$ (dof) & $W$ (m\AA)&&
 $log N$  & $b(\kms)$ & $\chi^2_r$ (dof) &  $W$ (m\AA) }
\startdata
{\ion{C}{2}}\tablenotemark{a} & 1036 &
              $13.2\pm 0.4 $  & $400\pm300$ & 1.1(22) &$91\pm8$&&
              $13.3\pm0.2$ & $300\pm200$& 1.3 (22) & $103\pm6$ \\
{\ion{C}{2}} & 1335 &
              $13.07\pm0.04$   & $260\pm 30$  &1.0 (9) &$22\pm2$&&
               $13.33\pm0.02$&$320\pm20$  & 0.64 (9) & $38\pm2$\\
{\ion{C}{3}} & 977 & 
             $13.13\pm0.03$  & $170\pm20$ & 0.79 (14) &$79\pm9$&&
              $13.07\pm0.06$ & $250\pm40$ &1.4 (14) & $64\pm9$ \\
{\ion{C}{4}} & 1548 &
              $13.24\pm0.02$   & $160\pm10$  &  5.3 (8) &$66\pm 1$\tablenotemark{c}&&
               $13.43\pm0.02$&$180\pm10$   & 6.7 (8) & $100\pm 2$\tablenotemark{c}\\
{\ion{C}{4}} & 1551 &
              $13.35\pm0.01$   & $158\pm 6$  & 0.24 (5) &$20\pm1$\tablenotemark{c}&&
               $13.53\pm0.01$ & $179\pm9$  & 1.2 (5) & $50\pm1$\tablenotemark{c}\\
{\ion{N}{5}} & 1243 &
              $13.1\pm0.4$  & $150\pm20$& 0.87 (17) &$9\pm7$&&
               $13.2\pm0.1$& $210\pm80$ & 0.73 (17) & $14\pm8$\\
{\ion{O}{1}}\tablenotemark{b} & 1302 &
              $13.0\pm0.1$  & $210\pm70$ & 1.0 (19) &$16\pm2$&&
               $13.49\pm0.08$& $380\pm80$  & 0.85 (19) & $42\pm3$\\
{\ion{O}{6}} & 1032 &
             $13.81\pm0.05$   & $210\pm30$  & 1.6 (7) &$76\pm5$&&
             $13.80\pm0.05$ & $250\pm 40$ &2.9 (7)  &  $68\pm3$\\
{\ion{O}{6}}\tablenotemark{a} & 1038 &
              $14.0\pm0.1$    & $450\pm60$  &1.1(22) &$91\pm8$&&
               $14.1\pm0.3$ &$550\pm50$ & 1.3 (22) & $103\pm6$\\
{\ion{Mg}{2}} & 2796 &
              $12.3\pm0.2$    & $200\pm100$  & 1.7  (7) &$80\pm20$&&
              $12.6\pm0.1$ & $300\pm100$ & 2.5 (7) &$150\pm20$\\
{\ion{Mg}{2}} & 2804 &
              $12.\pm5.$   & $<600$  & 0.98 (8) &$50\pm20$&&
               $12.7\pm0.3$& $300\pm100$ & 0.80 (8) &$70\pm20$\\
{\ion{Al}{2}} & 1671 &
              $11.55\pm0.04$   & $330\pm40$  & 0.56 (11)&$14\pm1$&&
               $11.72\pm0.03$& $380\pm40$  & 0.69 (11) &$20\pm1$\\
{\ion{Si}{2}} & 1260 &
              $11.88\pm0.07$  & $200\pm50$ & 0.95 (14) &$10\pm3$&&
               $12.39\pm0.06$& $450\pm90$ & 1.5 (14) & $37\pm3$\\
{\ion{Si}{2}}\tablenotemark{b} &1304  &
              $12.8\pm0.1$  & $230 \pm70$  & 1.0 (19) &$16\pm2$&&
               $13.3\pm0.1$& $700\pm200$ & 0.85 (19)  & $42\pm3$ \\
{\ion{Si}{2}} & 1527 &
              $12.80\pm0.05$  & $450\pm60$  & 0.69 (16) &$16\pm2$&&
              $13.08\pm 0.05$ & $580\pm80$  &1.4  (16) & $27\pm2$\\
{\ion{Si}{4}} &1394 &
              $12.57\pm0.01$   & $140\pm8$  & 0.52 (6) &$31\pm1$&&
              $12.75\pm0.01$ & $174\pm7$   & 0.49 (6) & $46\pm1$ \\
{\ion{Si}{3}} & 1207 &
              $12.66\pm 0.02$   & $220\pm 20$ & 0.89 (10) &$69\pm4$&&
                $12.65\pm0.01$& $253\pm8$  & 0.99 (10)  &$66\pm3$\\
{\ion{Fe}{2}} & 2344 &
              $12.5\pm0.3$   &  $500\pm300$ & 1.0 (13) &$11\pm5$&&
              $12.9\pm0.3$ & $<1000$ & 1.3 (13)& $22\pm5$\\
{\ion{Fe}{2}} & 2383 &
              $12.35\pm0.09$    & $500\pm100$  & 0.76 (19) &$37\pm7$&&
               $12.46\pm0.08$ & $420\pm90$   & 1.3 (19)  &$49\pm 7$\\
\enddata
\label{linefits}
\tablenotetext{a} {\phantom{ }Lines fitted together as a blend and equivalent width measured together.}
\tablenotetext{b} {\phantom{ }Lines fitted together as a blend and equivalent width measured together.}
\tablenotetext{c} {\phantom{ }Partial blend so blended portion not included in equivalent width measurement.}
\end{deluxetable*}

\section{Conclusions}
\label{conclusions}

We have developed a technique for measuring the {\it composite spectrum of \lya\ forest
absorbers} and applied it to 13,279 quasar spectra from the SDSS.
From our stacked spectra, we measure equivalent widths of Lyman series lines up to 
$n=7$ and column densities of 19 metal lines.
We find metal lines from species previously detected in the \lya\ forest
(\OVI, \NV, \CIV, \CIII, \SiIV\ and \SiIII), from new ionization states of oxygen 
(\OI), silicon (\SiII), and carbon (\CII), and from new elements magnesium (\MgII),
aluminum (\AlII) and iron (\FeII) not previously identified in high-redshift \lya\ forest studies.
Consistent results from the median and arithmetic mean stacked spectra
and from multiple lines of the same ionic species show that the
measured column densities are robust.
High $b$-parameters of the fitted metal lines provide a suggestive 
signal of large-scale clustering or high velocity dispersions in the
environment of the absorbers.

Analysis of the Lyman series implies typical column densities 
$N_{\rm HI} \approx 10^{15.4}\cm^{-2}$ for systems in our median stack.
The high-ionization metal-line species can then be explained assuming
typical physical conditions for the diffuse IGM and metallicity
[Fe/H]$\approx -2$, but reproducing the low-ionization species
requires much higher metallicities and higher densities, probably arising in
a minority sub-population of the absorption systems.

The composite spectrum technique introduced here has the potential
to teach us a great deal about physical conditions and enrichment
of the IGM and the spectral shape of the ionizing background radiation.
A natural next step is to compare composite spectra in narrower
bins of redshift and \lya\ optical depth, and to apply the technique
to the weaker but more abundant \lya\ forest absorbers that dominate
the mean opacity.  In this regard, we note that the high-redshift
quasar sample from the Baryon Oscillation Spectroscopic Survey of
SDSS-III \citep{schlegel09} will eventually exceed the size of the
current SDSS sample by an order of magnitude.
This technique may also be usefully applied to high-resolution spectra
and to {\it Hubble Space Telescope} spectra of the low-redshift
\lya\ forest, using the power of large numbers to bring out
features that are too weak to appear in even the best single-quasar
spectra.

\acknowledgments

We thank Jason X. Prochaska and Richard Pogge for useful discussions, and Benjamin 
Oppenheimer for the use of his CLOUDY output tables.
MP is supported in part by the Center for Cosmology and Astro-Particle
Physics at Ohio State University.
DW gratefully acknowledges the support of an AMIAS membership at
the Institute for Advanced Study.

Funding for the SDSS
and SDSS-II has been provided by the Alfred P. Sloan Foundation,
the Participating Institutions, the NSF, the US Department
of Energy, the National Aeronautics and Space Administration
(NASA), the Japanese Monbukagakusho, the Max Planck Society,
and the Higher Education Funding Council for England.
The SDSS Web site is at http://www.sdss.org/.

%

\clearpage


%

%



\end{document}